\documentclass[a4paper,11pt]{article}
\pdfoutput=1 
\usepackage{jheppub} 
\usepackage[T1]{fontenc} 
\usepackage[utf8]{inputenc}
\usepackage{amsfonts} 
\usepackage{graphics}
\usepackage{epsfig} 
\usepackage[english]{babel} 
\usepackage{caption} 
\usepackage{subcaption} 
\usepackage{amsmath}
\usepackage{bm}
\usepackage{multirow}
\usepackage{array,booktabs}
\usepackage{calc}
\usepackage[table]{xcolor}
\usepackage{color}  
\usepackage{makecell} 

\definecolor{aqua}{rgb}{0.0, 1.0, 1.0}
\definecolor{babyblue}{rgb}{0.54, 0.81, 0.94}
\definecolor{beaublue}{rgb}{0.74, 0.83, 0.9}
\definecolor{blizzardblue}{rgb}{0.93, 0.93, 0.93} 
\definecolor{cyan}{rgb}{0.0, 1.0, 1.0}
\newcolumntype{?}{!{\vrule width 0.8pt}}

\def\tanb{\tan\beta}
\def\hpm{H^\pm}
\def\mhpm{m_{\hpm}}
\def\mh{m_h}
\def\mH{m_H}
\def\ma{m_A}
\def\tb{\tan\beta}

\def\cb{\cot\beta}
\def\bma{\beta-\alpha}  

\newcommand{\RN}[1]{
  \textup{\uppercase\expandafter{\romannumeral#1}}%
\renewcommand\thesubfigure{(\alph{subfigure})} 
\captionsetup[sub]{labelformat=simple} 
}
 
\title{\boldmath Observability of 2HDM neutral Higgs bosons with different masses at future $e^+e^-$ linear colliders} 
\author[]{Majid Hashemi and Gholamhossein Haghighat}
\affiliation[]{Physics Department, College of Sciences, Shiraz University, \\ Shiraz, 71946-84795, Iran}
\emailAdd{hashemi$\_$mj$@$shirazu.ac.ir, hosseinhaqiqat$@$gmail.com} 

\abstract{Assuming two Higgs doublet model (2HDM) at SM-like scenario as the theoretical framework, this study addresses the question of observability of heavy neutral CP-even and CP-odd Higgs bosons $H$ and $A$ at a linear collider operating at $\sqrt{s}=1$ TeV. The signal production channel is $e^- e^+ \rightarrow A H \rightarrow ZHH$ with subsequent leptonic decay of the $Z$ boson and Higgs bosons decays into $b$ quark pairs. Therefore, to be specific, type-I 2HDM is used to allow dominant Higgs boson decay to b-quark pairs below the top quark pair production threshold. Two benchmark points with mass ranges $150\leq m_H\leq 200$ and $250\leq m_A\leq 300$ are simulated. The relevant energy and momentum smearing is applied and appropriate selection cuts are imposed to enrich signal events. Results indicate that both Higgs bosons are observable with signals exceeding $5\sigma$ significance with possibility of mass measurement at the integrated luminosity of 500 $fb^{-1}$.  }       

\begin{document}
\maketitle 
\flushbottom
 
\section{Introduction}
The standard model (SM) of elementary particles has been verified by many experiments and played an important role in understanding a wide range of phenomena. Existence of the Higgs boson \cite{Englert1,Higgs1,Higgs2,Kibble1,Higgs3,Kibble2} as one of the most important predictions of the standard model was verified experimentally \cite{HiggsObservationCMS,HiggsObservationATLAS} and triggered an increasing interest in studying SM extensions. Extensions of the SM are also motivated by supersymmetry \cite{MSSM1}, axion models \cite{KIM1}, the SM inability to explain the universe baryon asymmetry \cite{Trodden}, etc. The standard model uses the simplest possible scalar structure with one $SU(2)$ Higgs doublet. Such an assumption leads to the prediction of a single Higgs boson. However, employing two $SU(2)$ Higgs doublets leads to a kind of SM extension which may resolve the present unsolved problems of the physics. 

As one of the simplest extensions of the standard model, two-Higgs-doublet model (2HDM) \cite{2hdm_TheoryPheno,2hdm1,2hdm2,2hdm3,2hdm4_CompositeHiggs,2hdm_HiggsSector1,2hdm_HiggsSector2,Campos:2017dgc} has emerged as an important candidate which predicts five Higgs bosons, four of which are assumed to be, yet undiscovered, Higgs bosons and the fifth one (the lightest one) is assumed to be the same as the observed SM Higgs boson. Prediction of the existence of five Higgs bosons is a direct consequence of assuming two $SU(2)$ Higgs doublets in this model. Two out of the four undiscovered Higgs bosons are neutral scalar and pseudoscalar Higgs bosons $H$ and $A$, and the others are charged Higgs bosons $H^\pm$. This paper focuses on the neutral Higgs bosons $H$ and $A$ in the context of the Type-\RN{1} 2HDM at SM-like scenario and addresses the question of observability of these Higgs bosons at a linear collider operating at the center-of-mass energy of $\sqrt{s}=1$ TeV. Type-\RN{1} 2HDM is one of the four types of the 2HDM which naturally conserve flavor and are derived from imposing the discrete $Z_2$ symmetry. 

In this study, the process $e^- e^+ \rightarrow A H\rightarrow ZHH \rightarrow \ell\bar{\ell} b\bar{b}b\bar{b}$, where $\ell\bar{\ell}$ is an electron or a muon pair, is assumed as the signal process. The decay mode $H\rightarrow b\bar{b}$ is motivated at high $\tan\beta$ by the significant enhancement compared to other modes. This is due to the fact that Higgs-fermion couplings are proportional to the same $\cb$ factor which is canceled out when branching ratios of Higgs boson decays are calculated.  Therefore as long as the Higgs boson mass is below the on-shell production of the top quark pair production, $H \to b\bar{b}$ remains dominant. Despite having a small branching ratio ($\approx 0.066$), the decay mode $Z\rightarrow e^-e^+$ or $\mu^-\mu^+$ is chosen for signal to benefit from the clear signature electrons and muons provide at linear colliders. 

To investigate the observability of the signal, two benchmark points in the parameter space of the Type-\RN{1} 2HDM are assumed to search for heavy neutral CP-even and CP-odd Higgs bosons $H$ and $A$ with the help of appropriate selection cuts. We finally try to reconstruct masses of the Higgs bosons. Results indicate that, for both benchmark points, the Higgs bosons $H$ and $A$ are observable at a 1 TeV collider with $5\sigma$ signals at integrated luminosities 46 and 112 $fb^{-1}$ respectively.  

In what follows, we provide a brief introduction to the 2HDM and its flavor conserving types, and then the signal and background analysis and results will be provided.

\section{Two-Higgs-doublet model} 
A general 2HDM assumes the Higgs potential to be
\begin{equation}
  \begin{aligned}
    \mathcal{V} = &m_{11}^2\Phi_1^\dagger\Phi_1+m_{22}^2\Phi_2^\dagger\Phi_2
    -\left[m_{12}^2\Phi_1^\dagger\Phi_2+\mathrm{h.c.}\right]
    \\
    &+\frac{1}{2}\lambda_1\left(\Phi_1^\dagger\Phi_1\right)^2
    +\frac{1}{2}\lambda_2\left(\Phi_2^\dagger\Phi_2\right)^2
    +\lambda_3\left(\Phi_1^\dagger\Phi_1\right)\left(\Phi_2^\dagger\Phi_2\right)
    +\lambda_4\left(\Phi_1^\dagger\Phi_2\right)\left(\Phi_2^\dagger\Phi_1\right)
    \\&+\left\{
    \frac{1}{2}\lambda_5\left(\Phi_1^\dagger\Phi_2\right)^2
    +\left[\lambda_6\left(\Phi_1^\dagger\Phi_1\right)
      +\lambda_7\left(\Phi_2^\dagger\Phi_2\right)
      \right]\left(\Phi_1^\dagger\Phi_2\right)
    +\mathrm{h.c.}\right\},
  \end{aligned}
  \label{lag}
\end{equation}
where $\Phi_1$ and $\Phi_2$ are $SU(2)$ Higgs doublets. Employing the extended scalar structure with two Higgs doublets leads to the prediction of three neutral Higgs bosons $h$, $H$ and $A$, and two charged Higgs bosons $H^\pm$. $h$ and $H$ are scalar CP-even bosons and $A$ is a pseudoscalar CP-odd boson. Working in the ``physical basis'', physical Higgs masses, $\tanb$, CP-even Higgs mixing angle $\alpha$, $m_{12}^2$, $\lambda_6$ and $\lambda_7$ are parameters of the model and must be determined \cite{2hdm_TheoryPheno}. $m_{11}^2$ and $m_{22}^2$ in the Higgs potential \ref{lag} are determined by the minimization conditions for a minimum of the vacuum once $\tb$ is determined. Imposing discrete $Z_2$ symmetry \cite{2hdm2,2hdm3,2hdm4_CompositeHiggs} to avoid tree level flavor-changing neutral currents (FCNC) implies that the values of the parameters $\lambda_6$, $\lambda_7$ and $m_{12}^2$ must be zero. However, setting $\lambda_6$, $\lambda_7$ to zero and allowing a non-zero value for $m_{12}^2$, $Z_2$ symmetry is softly broken in 2HDM. Imposing $Z_2$ symmetry restricts Higgs coupling to fermions and implies that there are four types of the 2HDM which naturally conserve flavour. Table \ref{coupling} shows how Higgs doublets couple to fermions in different types.
\begin{table}[h]
\normalsize
\fontsize{11}{7.2} 
    \begin{center}
         \begin{tabular}{ >{\centering\arraybackslash}m{.8in} >{\centering\arraybackslash}m{.4in} >{\centering\arraybackslash}m{.4in} >{\centering\arraybackslash}m{.4in}   }
& {$u_R^i$} & {$d_R^i$} & {$\ell_R^i$} \parbox{0pt}{\rule{0pt}{1ex+\baselineskip}}\\ \Xhline{3\arrayrulewidth}
 {Type \RN{1}} &$\Phi_2$ &$\Phi_2$ &$\Phi_2$ \parbox{0pt}{\rule{0pt}{1ex+\baselineskip}}\\ 
 {Type $\RN{2}$} &$\Phi_2$ &$\Phi_1$ &$\Phi_1$ \parbox{0pt}{\rule{0pt}{1ex+\baselineskip}}\\ 
{Type X} &$\Phi_2$ &$\Phi_2$ &$\Phi_1$  \parbox{0pt}{\rule{0pt}{1ex+\baselineskip}}\\ 
{Type Y} &$\Phi_2$ &$\Phi_1$ &$\Phi_2$  \parbox{0pt}{\rule{0pt}{1ex+\baselineskip}}\\ \Xhline{3\arrayrulewidth}
  \end{tabular}
\caption{Higgs coupling to up-type quarks, down-type quarks and leptons in different types. The superscript $i$ is a generation index.  \label{coupling}}
  \end{center}
\end{table}
The types ``X'' and ``Y'' are also called ``lepton-specific'' and ``flipped'' respectively. Applying the coupling prescription of table \ref{coupling}, Higgs-fermion interaction part of the Lagrangian becomes \cite{2hdm_TheoryPheno}
\begin{equation}
\begin{aligned}
\mathcal{L}_{\ Yukawa}\ =\ & -\ \sum_{f=u,d,\ell}\ \dfrac{m_f}{v}\ \Big(\xi_{h}^{f}\bar{f}fh\ +\ \xi_{H}^{f}\bar{f}fH\ -\ i\xi_{A}^{f}\bar{f}\gamma_5fA \Big)\\
&-\ \Bigg\{\dfrac{\sqrt{2}V_{ud}}{v}\bar{u}\ \big(m_u\xi_A^uP_L\ +\ m_d\xi_A^dP_R\big)\ dH^+\ +\ \dfrac{\sqrt{2}m_\ell\xi_A^\ell}{v}\overline{\nu_L}\ell_RH^+\ +\ H.c. \Bigg\}.
\label{yukawa1}
\end{aligned}
\end{equation}
Table \ref{xi} provides $\xi^X_Y$ factors corresponding to different types.
\begin{table}[h]
\normalsize
\fontsize{11}{7.2} 
    \begin{center}
         \begin{tabular}{ >{\centering\arraybackslash}m{.5in} >{\centering\arraybackslash}m{.55in}  >{\centering\arraybackslash}m{.55in} >{\centering\arraybackslash}m{.55in}  >{\centering\arraybackslash}m{.55in}  }
& {\RN{1}} & {$\RN{2}$} & {X} & {Y} \parbox{0pt}{\rule{0pt}{1ex+\baselineskip}}\\ \Xhline{3\arrayrulewidth}
 {$\xi_h^u$} &$c_\alpha/s_\beta$ &$c_\alpha/s_\beta$ &$c_\alpha/s_\beta$  &$c_\alpha/s_\beta$  \parbox{0pt}{\rule{0pt}{1ex+\baselineskip}}\\ 
 {$\xi_h^d$} &$c_\alpha/s_\beta$ &$-s_\alpha/c_\beta$ &$c_\alpha/s_\beta$  &$-s_\alpha/c_\beta$  \parbox{0pt}{\rule{0pt}{1ex+\baselineskip}}\\ 
 {$\xi_h^\ell$} &$c_\alpha/s_\beta$ &$-s_\alpha/c_\beta$ &$-s_\alpha/c_\beta$   &$c_\alpha/s_\beta$  \parbox{0pt}{\rule{0pt}{1ex+\baselineskip}}\\ 
 {$\xi_H^u$} &$s_\alpha/s_\beta$ &$s_\alpha/s_\beta$ &$s_\alpha/s_\beta$  &$s_\alpha/s_\beta$  \parbox{0pt}{\rule{0pt}{1ex+\baselineskip}}\\ 
 {$\xi_H^d$} &$s_\alpha/s_\beta$ &$c_\alpha/c_\beta$ &$s_\alpha/s_\beta$  &$c_\alpha/c_\beta$  \parbox{0pt}{\rule{0pt}{1ex+\baselineskip}}\\ 
 {$\xi_H^\ell$} &$s_\alpha/s_\beta$ &$c_\alpha/c_\beta$ &$c_\alpha/c_\beta$  &$s_\alpha/s_\beta$  \parbox{0pt}{\rule{0pt}{1ex+\baselineskip}}\\ 
 {$\xi_A^u$} &$\cot\beta$ &$\cot\beta$ &$\cot\beta$  &$\cot\beta$  \parbox{0pt}{\rule{0pt}{1ex+\baselineskip}}\\ 
 {$\xi_A^d$} &$-\cot\beta$ &$\tan\beta$ &$-\cot\beta$  & $\tan\beta$ \parbox{0pt}{\rule{0pt}{1ex+\baselineskip}}\\ 
 {$\xi_A^\ell$} &$-\cot\beta$ &$\tan\beta$ &$\tan\beta$  & $-\cot\beta$ \parbox{0pt}{\rule{0pt}{1ex+\baselineskip}}\\ \Xhline{3\arrayrulewidth}
  \end{tabular}
\caption{$\xi^X_Y$ factors corresponding to different types ($c_x\equiv\cos x$ and $s_x\equiv\sin x$). \label{xi}}
  \end{center}
\end{table}
In order to respect experimental observations, it is assumed that the lightest Higgs boson $h$ predicted by the 2HDM is the same as the discovered SM Higgs boson and thus the SM-like scenario is chosen by assuming $\sin(\bma)=1$ \cite{2hdm_TheoryPheno}. Therefore the $h$-fermion couplings of the Yukawa Lagrangian of the 2HDM reduce to the corresponding couplings of the standard model. As a result, the neutral Higgs part of the Yukawa Lagrangian takes the form \cite{Barger_2hdmTypes}
\begin{equation}
\begin{aligned}
\mathcal{L}_{\ Yukawa}\ =\ & -v^{-1}  \Big(\ m_d\ \bar{d}d\ +\ m_u\ \bar{u}u\ +\ m_\ell\ \bar{\ell}\ell\ \Big)\ h \\
      & +v^{-1} \Big(\ \rho^dm_d\ \bar{d}d\ +\ \rho^um_u\ \bar{u}u\ +\ \rho^\ell m_\ell\ \bar{\ell}\ell\ \Big)\ H \\
& +iv^{-1}\Big(-\rho^dm_d\ \bar{d}\gamma_5d\ +\ \rho^um_u\ \bar{u}\gamma_5u\ -\ \rho^\ell m_\ell\ \bar{\ell}\gamma_5\ell\ \Big)\ A,
\label{yukawa2}
\end{aligned}
\end{equation}
 where $\rho^X$ factors are given in table \ref{rho}.
\begin{table}[h]
\normalsize
\fontsize{11}{7.2} 
    \begin{center}
         \begin{tabular}{ >{\centering\arraybackslash}m{.5in}  >{\centering\arraybackslash}m{.55in}  >{\centering\arraybackslash}m{.55in} >{\centering\arraybackslash}m{.55in}  >{\centering\arraybackslash}m{.55in}}
& {\RN{1}} & {$\RN{2}$} & {X} & {Y} \parbox{0pt}{\rule{0pt}{1ex+\baselineskip}}\\ \Xhline{3\arrayrulewidth}
 {$\rho^d$} &$\cot{\beta}$ &$- \tan\beta$ &$\cot\beta$ &$-\tan\beta$ \parbox{0pt}{\rule{0pt}{1ex+\baselineskip}}\\ 
{$\rho^u$} &$\cot{\beta}$ &$\cot\beta$ &$\cot\beta$ &$\cot\beta$  \parbox{0pt}{\rule{0pt}{1ex+\baselineskip}}\\   
{$\rho^\ell$} &$\cot{\beta}$ &$- \tan\beta$ &$-\tan\beta$&$\cot\beta$  \parbox{0pt}{\rule{0pt}{1ex+\baselineskip}}\\ \Xhline{3\arrayrulewidth} 
 \end{tabular}
\caption{$\rho^X$ factors of the neutral Higgs part of the Yukawa Lagrangian corresponding to different types. \label{rho}}
  \end{center}    
\end{table}  
Different types of the 2HDM show different characteristics \cite{2hdm_HiggsSector2} due to the difference among the factors. As table \ref{rho} shows, factors corresponding to the Type-\RN{1} 2HDM increase as $\tb$ decreases. Such a behaviour is one of the motivations behind working in low $\tb$ regime in the context of this type. 

\section{Signal process}
The signal process is assumed to be $e^- e^+ \rightarrow A H$ in the context of the 2HDM Type-\RN{1}. The Higgs bosons are selected with different masses to provide possibility of $A \rightarrow ZH$ decay. Scenarios with equal masses were studied earlier leading to promising results under the same collider conditions \cite{Hbb}. The two scalar Higgs bosons then decay like $H \rightarrow b\bar{b}$ which is dominated in 2HDM Type-\RN{1} and the $Z$ boson undergoes $Z \rightarrow \ell\bar{\ell}$ decay where $\ell\bar{\ell}$ is an electron or a muon pair and $b$ is the bottom quark. The center-of-mass energy of $\sqrt{s}=1$ TeV is assumed for the initial collision at a linear collider. Two benchmark points with different mass hypotheses are assumed as shown in table \ref{BPs}. 
\begin {table}[h]  
\begin{center}
\begin{tabular}{cccc} 
& \multicolumn{1}{ c }{BP1} & BP2 \\ \Xhline{3\arrayrulewidth}
\multicolumn{1}{ c  }{$m_{h}$} & \multicolumn{2}{ c }{125} \\ 
\multicolumn{1}{ c  }{$m_{H}$} & 150 & 200 \\
\multicolumn{1}{ c  }{$m_{A}$} & 250 & 300  \\ 
\multicolumn{1}{ c  }{$m_{H^\pm}$} & 250 & 300  \\ 
\multicolumn{1}{ c  }{$m_{12}^2$} & 2001-2223 & 3722-3972  \\  
\multicolumn{1}{ c  }{$\tan\beta$} & \multicolumn{2}{ c }{10} \\ 
\multicolumn{1}{ c  }{$\sin(\beta-\alpha)$} & \multicolumn{2}{ c }{1} \\ \Xhline{3\arrayrulewidth}
\end{tabular}
\caption {Assumed benchmark points. $\mH,\mh,\ma,\mhpm$ are physical masses of the Higgs bosons. The $m^2_{12}$ range satisfying theoretical requirements is provided for each scenario. \label{BPs}}
\end{center}   
\end {table}
The physical mass of the $H$ Higgs boson is assumed to take values $150$ and $200$ GeV, and the mass splitting between $H$ and $A$ Higgs bosons is assumed to be 100 GeV for on-shell $Z$ boson production. The chosen Higgs boson masses are checked to be consistent with results of 86 analyses with the use of \texttt{HiggsBounds 4.3.1} \cite{HB} and \texttt{HiggsSignals 1.3.0} \cite{HS}. The value of $\tan\beta$ is also set to 10 for both benchmark points. For each scenario there is a range of $m^2_{12}$ parameter (quoted in Tab. \ref{BPs}) which satisfies theoretical requirements of potential stability \cite{Deshpande}, perturbativity and unitarity \cite{Huffel,Maalampi,KANEMURA,GAKEROYD} which are all checked using \texttt{2HDMC 1.7.0} \cite{2hdmc1,2hdmc2}.

Masses of the Higgs bosons $A$ and $H^\pm$ are assumed to be equal for both of the benchmark points to make sure that the experimental constraint \cite{BERTOLINI,DENNER} is satisfied. This experimental constraint resulted from the measurement performed at LEP \cite{Yao} and puts a limit on the deviation of the parameter $\rho=m_W^2(m_Z\cos\theta_W)^{-2}$ from its standard model value. Since it is demonstrated that the deviation of this parameter is negligible if any of the conditions \cite{drho,Gerard:2007kn} 
\begin{equation}
m_A=m_{H^\pm},\,\,\, m_H=m_{H^\pm},
\label{negligibledrho}
\end{equation}
is satisfied, the assumed benchmark points are guaranteed to be consistent with the mentioned experimental constraint. Flavor physics data constrains charged Higgs mass by the limit $m_{H^\pm}>480$ GeV in the Type-\RN{2} and Type-Y 2HDM \cite{Misiak,Misiak2017}. However, since the charged Higgs coupling to quarks in the Type-\RN{1} 2HDM depends on $\cb$ and differs from the same coupling in the Type-\RN{2} and Type-Y 2HDM, the mentioned constraint does not limit choice of charged Higgs mass in this study. The assumed benchmark points are totally consistent with the results of ATLAS direct investigation \cite{Atlas2HDMconstraints} on 2HDM. Moreover, as indicated in \cite{lep1,lep2,lepexclusion2}, the limits $m_A\geq93.4$ GeV and $m_{H^\pm}\geq78.6$ GeV are already satisfied by the current analysis. The LHC experiments have recently excluded the region $m_{A/H}=200-400$ GeV for $\tb\geq5$ \cite{CMSNeutralHiggs,ATLASNeutralHiggs}. However, since the Type-\RN{1} 2HDM differs considerably from the MSSM in structure, mass hypotheses in this study are not required to satisfy these conditions. Therefore, it can be concluded that the assumed benchmark points satisfy all of the theoretical and experimental constraints and can be used to generate signal events. 

According to the full Lagrangian of the Type-\RN{1} 2HDM, the $Z$-$H$-$A$ vertex depends on $\sin(\bma)$ which is assumed to be unity in the SM-like scenario. This vertex appears both in the production process, i.e., $e^+e^- \rightarrow Z^* \rightarrow HA$ and the subsequent decay $A\rightarrow ZH$. Therefore the production process followed by $A \rightarrow ZH$ decay is independent of $\tan\beta$ as long as $\sin(\bma)=1$. We obtain BR($A \rightarrow ZH$)$\simeq 0.998$ and BR($H \rightarrow b\bar{b}$)$\simeq 0.71$ for the two benchmark scenarios using \texttt{2HDMC 1.7.0}.  
According to the signal process, the produced $Z$ boson annihilates into a lepton pair ($\mu^-\mu^+$ or $e^-e^+$). Branching ratio of this decay mode ($\approx 0.066$) is so small compared with the hadronic decay mode. Despite this fact, the leptonic decay is chosen since leptons provide a simple and clear signature at linear colliders and this feature can compensate for the smallness of the branching ratio. Each signal event results in two $H$ Higgs bosons which are assumed to decay into $b$ quark pairs. The resulting $b$ quarks annihilate into hadronic jets which are used to reconstruct the Higgs boson. Reconstruction of the $H$ and $Z$ bosons is then followed by $A$ reconstruction as fully discussed in the following sections.

Table \ref{sXsec} shows signal process cross section corresponding to the assumed benchmark points obtained by \texttt{PYTHIA 8.2.15} \cite{pythia82}. 
\begin {table}[h]
\begin{center}
\begin{tabular}{ccc}
& \multicolumn{1}{ c }{BP1} & BP2 \\ \Xhline{3\arrayrulewidth}
\multicolumn{1}{ c  }{Signal cross section [fb]} & 0.338 & 0.207 \\ \Xhline{3\arrayrulewidth}
\end{tabular}
\caption {Cross section of the signal process assuming different benchmark points. \label{sXsec}}
\end{center}
\end {table}
As seen, the benchmark point with heavier Higgs masses corresponds to the smaller cross section. Therefore, observing the heavier Higgs boson is expected to be more difficult. Background processes contributing to this analysis include top quark pair production, $W^\pm$ pair production, $Z$ pair production and $Z/\gamma$ production. Cross sections corresponding to the background processes are also obtained by \texttt{PYTHIA 8.2.15} and are provided in Tab. \ref{bgXsec}.
\begin {table}[h]
\begin{center}
\begin{tabular}{ccccc}
& \multicolumn{1}{ c }{$t\bar{t}$} & $W^+W^-$ & $ZZ$ & $Z/\gamma$ \\ \Xhline{3\arrayrulewidth}
\multicolumn{1}{ c  }{Cross section [fb]} & 211.1 & 3163 & 234.7 & 4335 \\ \Xhline{3\arrayrulewidth}
\end{tabular}
\caption {Background cross sections. \label{bgXsec}}
\end{center}
\end {table}

\section{Analysis} 
\begin{figure}[h!]
  \centering
    \begin{subfigure}[b]{0.59\textwidth}
    \centering
    \includegraphics[width=\textwidth]{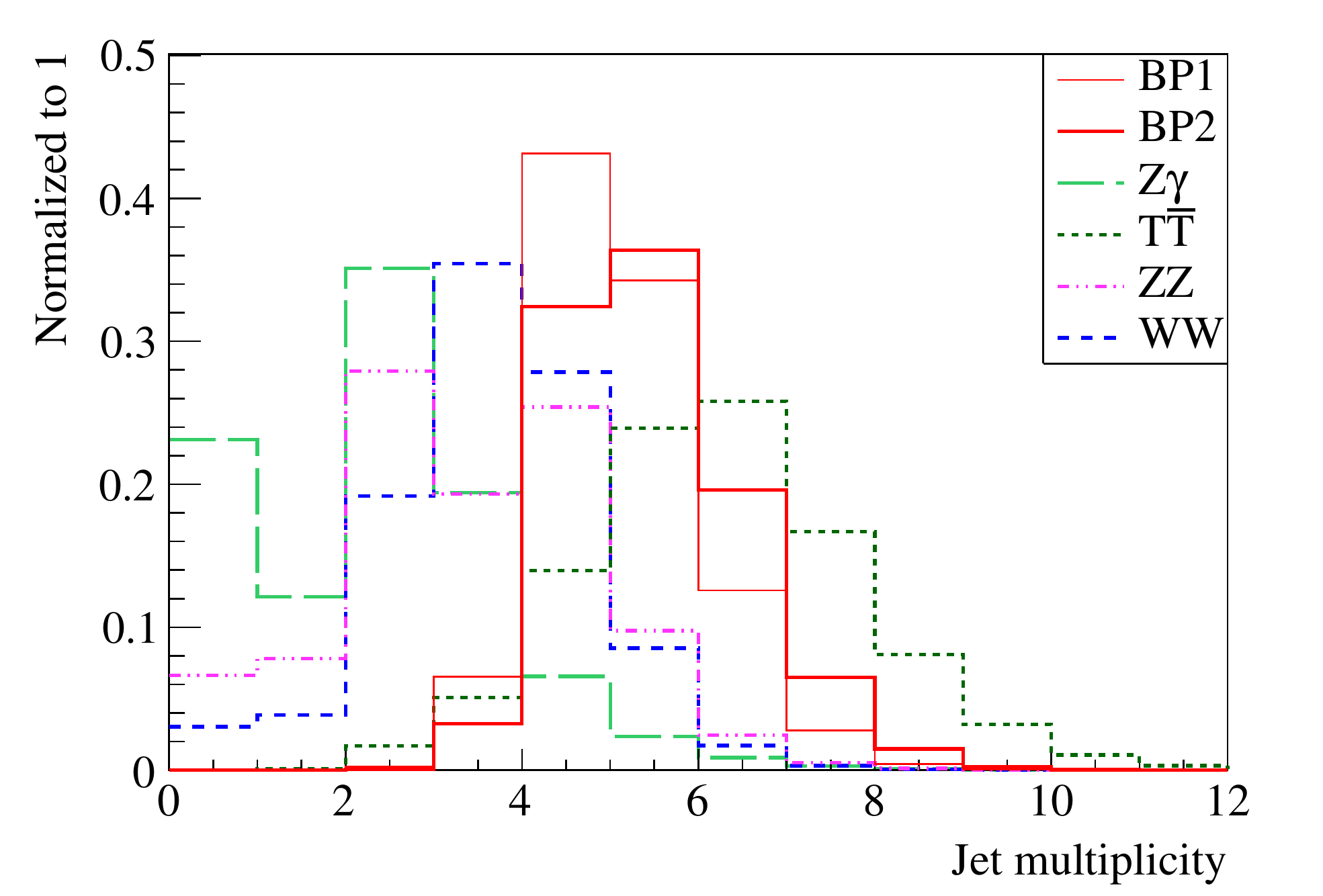}
    \caption{}
    \label{hnjets} 
    \end{subfigure}  
        \quad    
    \begin{subfigure}[b]{0.59\textwidth}
    \centering
    \includegraphics[width=\textwidth]{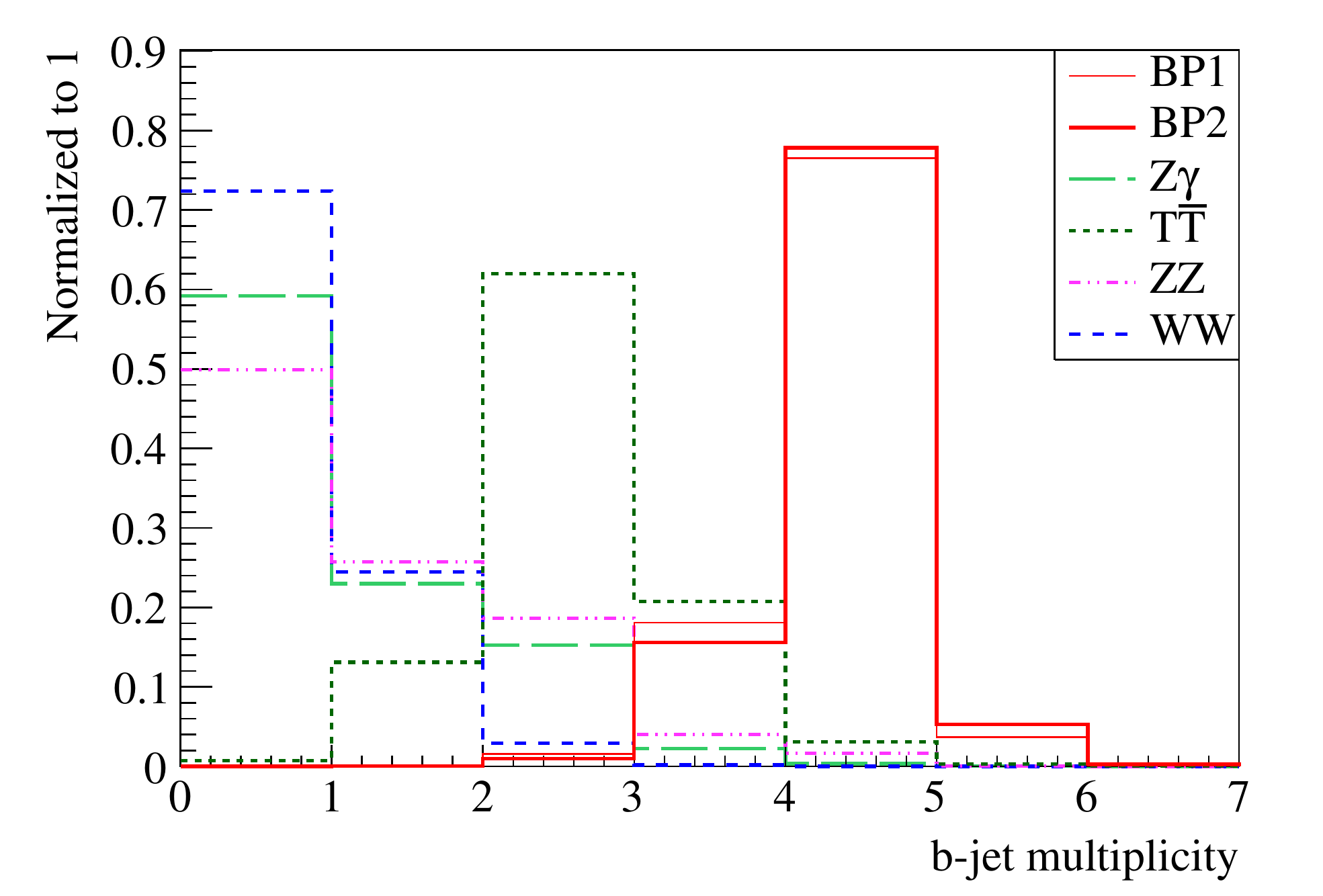}
    \caption{}
    \label{hnbjets}
    \end{subfigure}
  \caption{a) Jet multiplicity and b) $b$-jet multiplicity obtained for signal and background processes.}
  \label{hnjetshnbjets} 
\end{figure}

To generate signal events, basic parameters of the model are generated in SLHA (SUSY Les Houches Accord) format by \texttt{2HDMC 1.7.0} and the output is passed to \texttt{PYTHIA 8.2.15} for event generation. Background events are also generated by \texttt{PYTHIA 8.2.15}. Based on the characteristics of the signal and background events, appropriate event selection cuts are applied to enrich signal events. FASTJET 3.1.0 \cite{fastjet1,fastjet2} is used to perform jet reconstruction. According to properties of the jets, anti-$k_t$ algorithm \cite{antikt} with the cone size $\Delta R=\sqrt{(\Delta\eta)^2+(\Delta\phi)^2}=0.4$ is employed. Here, $\eta=-\textnormal{ln}\tan(\theta/2)$ and $\phi$ ($\theta$) is the azimuthal (polar) angle with respect to the beam axis. After jets are identified, jet energy smearing is applied to jets according to energy resolution $\sigma/E=3.5\, \%$ \cite{cliccdr}. Jets are required to satisfy the conditions
\begin{equation}
\bm{{p_T}}_{\bm{jet}}\geq10\ GeV,\ \ \  \vert \bm{\eta}_{\bm{jet}} \vert \leq 4,
\label{jetconditions}
\end{equation}
where $p_T$ is the transverse momentum. Counting jets in each event results in jet multiplicity distributions as shown in Fig. \ref{hnjets}.
Based on the obtained distributions, the condition
\begin{equation}
\bm{N}_{\textbf{\emph{jet}}} \geq 4,
\label{jetsnumbercondition} 
\end{equation}
is applied, where $N_{\textbf{\emph{jet}}}$ is the number of jets. Applying the $b$-tagging algorithm to jets, the $b$-jet multiplicity distributions are obtained as shown in Fig. \ref{hnbjets}. The $b$-tagging method is based on $b$-tag efficiency, $c$-jets mis-tag rate and light jets mis-tag rate assumed to be $0.7, 0.07$ and $0.003$ respectively \cite{cliccdr}. Based on the distributions of Fig. \ref{hnbjets}, the selection cut 
\begin{equation}
\bm{N}_{\textbf{\emph{b-jet}}} \geq 3.
\label{bjetsnumbercondition} 
\end{equation}
is applied to include events with at least three $b$-jets.

\begin{figure}[h]
  \centering
  \includegraphics[width=0.59\textwidth]{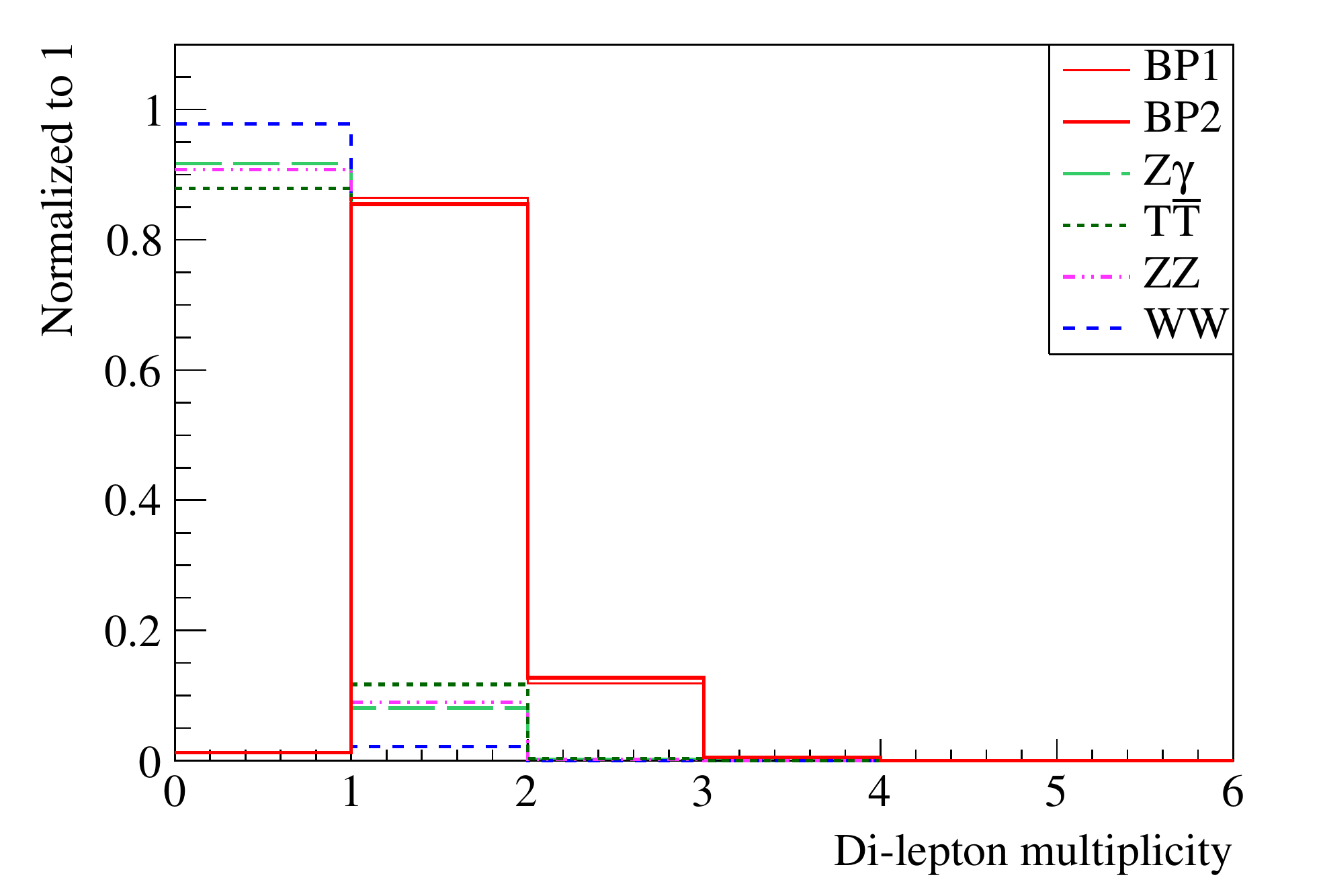}
  \caption{Di-lepton multiplicity distributions of the signal and background events. }
\label{hndileptons}
\end{figure}

Leptons (electrons and muons) present in the events are identified and momentum smearing according to the momentum resolution $\sigma_{p_T}/{p_T^2} = 2 \times 10^{-5}$ GeV$^{-1}$ \cite{cliccdr} is applied to them. Counting the number of electrons and muons which satisfy the conditions 
\begin{equation}
{\bm{p_T}}_{\bm{e,\mu}}\geq5\ GeV,\ \ \  \vert \bm{\eta}_{\bm{e,\mu}} \vert \leq 4, 
\label{leptonsconditions}
\end{equation}
the number of di-leptons is obtained. A di-lepton can be a di-electron or a di-muon. Figure \ref{hndileptons} shows the obtained di-lepton multiplicity corresponding to different processes. 
The selection cut
\begin{equation}
\bm{N}_{\bm{\ell\bar{\ell}}} \geq1, 
\label{dileptonscondition}
\end{equation} 
where $N_{\ell\bar{\ell}}$ is the number of di-leptons, is now applied to rule out events with no di-lepton.

Events with at least one di-lepton are then subject to conditions
\begin{equation}
60 \leq \bm{I. M.}_{\, \bm{\ell\bar{\ell}}}  \leq 100\ GeV,
\label{leptonsconditions}
\end{equation}
where $I. M._{\, {\ell\bar{\ell}}}$ is the invariant mass of the di-lepton. An event is selected if it has a di-lepton satisfying these conditions. 

Analyzing the $b$-tagged jets, decay products of the $H$ Higgs bosons are identified. In each event, the parameter 
\begin{equation} 
\Delta R=\sqrt{(\Delta\eta)^2+(\Delta\phi)^2},
\label{deltaR}
\end{equation}
is computed for all possible $b$-jet pairs and the pair for which the $\Delta R$ value is minimized, is identified as the true pair which originates from the $H$ Higgs boson. Computing the invariant masses of the identified true $b$-jet pairs, $H$ Higgs mass is reconstructed and a mass distribution is obtained as fully discussed later.

Having reconstructed the $H$ Higgs boson, another selection cut is applied to events and then the $A$ Higgs boson mass will be reconstructed. Events which satisfy any of the conditions 
\begin{equation}  
\begin{aligned}
& \bm{N}_{\bm{b\bar{b}}} \geq2,\\
& \bm{\Delta R}{({\bm{ZH}})} \leq 1\ \ (if\ {N}_{b\bar{b}}=1), 
\label{dibnumbercondition}
\end{aligned} 
\end{equation}
where $N_{b\bar{b}}$ is the number of identified true $b$-jet pairs, pass the selection cut. ${\Delta R}{({{ZH}})}$ is the spatial distance (defined by Eq. \ref{deltaR}) between the reconstructed $H$ boson (originating from the $b$-jet pair) and the reconstructed $Z$ boson (originating from the di-lepton). Therefore, events with two or more true $b$-jet pairs, and events with one true $b$-jet pair which satisfy ${\Delta R}{({ZH})} \leq 1$, survive the last selection cut.

In events with one $b$-jet pair, computing the invariant mass of the $b$-jet pair and the lepton pair reults in a value for the mass of A boson. In events with two $b$-jet pairs, the $b$-jet pair which its corresponding reconstructed $H$ boson has smaller spatial distance (see Eq. \ref{deltaR}) from the reconstructed $Z$ boson (originating from the di-lepton) is identified as the decay product of the $A$ Higgs boson. In such events also a value for the A boson mass is obtained as the invariant mass of the identified $b$-jet pair and the lepton pair. 

Event selection efficiencies corresponding to the applied selection cuts are provided in tables \ref{signalefftab} and \ref{backgroundefftab}. $H$ and $A$ candidate mass distributions are obtained after four and five cuts respectively. Total efficiencies corresponding to the first four cuts and all the five cuts are also provided. 
\begin{table}[h]
\normalsize
\fontsize{11}{7.2} 
    \begin{center}
         \begin{tabular}{ >{\centering\arraybackslash}m{1.3in}  >{\centering\arraybackslash}m{.45in}  >{\centering\arraybackslash}m{.45in} >{\centering\arraybackslash}m{.45in}  } 
  & {BP1} & {BP2}  \parbox{0pt}{\rule{0pt}{1ex+\baselineskip}}\\ \Xhline{3\arrayrulewidth}
    {$N_{jet}\geq4$} & 0.932 & 0.967  \parbox{0pt}{\rule{0pt}{1ex+\baselineskip}}\\ 
   {${N}_{{\emph{b-jet}}} \geq 3$} & 0.984 & 0.990  \parbox{0pt}{\rule{0pt}{1ex+\baselineskip}}\\ 
    {$N_{\ell\bar{\ell}}\geq1$} & 0.988 & 0.988   \parbox{0pt}{\rule{0pt}{1ex+\baselineskip}}\\ 
{$60 \leq {I. M.}_{\, {\ell\bar{\ell}}}  \leq 100$} & 0.880 & 0.864  \parbox{0pt}{\rule{0pt}{1ex+\baselineskip}}\\ 
    {\textbf{Total eff.}} & \textbf{0.797} & \textbf{0.817}  \parbox{0pt}{\rule{0pt}{1ex+\baselineskip}}\\  \Xhline{3\arrayrulewidth}
   {${\Delta R}{({{ZH}})}\leq1$} & 0.913 & 0.922  \parbox{0pt}{\rule{0pt}{1ex+\baselineskip}}\\ 
    {\textbf{Total eff.}} & \textbf{0.728} & \textbf{0.753}  \parbox{0pt}{\rule{0pt}{1ex+\baselineskip}}\\ \Xhline{3\arrayrulewidth}
        \end{tabular}
\caption{Event selection efficiencies corresponding to the signal process assuming two benchmark points. \label{signalefftab}}
  \end{center}
\end{table}
\begin{table}[h] 
\normalsize
\fontsize{11}{7.2} 
    \begin{center} 
         \begin{tabular}{ >{\centering\arraybackslash}m{1.3in}  >{\centering\arraybackslash}m{.62in}  >{\centering\arraybackslash}m{.62in} >{\centering\arraybackslash}m{.62in} >{\centering\arraybackslash}m{.62in}  >{\centering\arraybackslash}m{.62in} }
& {$t\bar{t}$} & {$WW$} & {$ZZ$} & {$Z\gamma$ } \parbox{0pt}{\rule{0pt}{1ex+\baselineskip}}\\ \Xhline{3\arrayrulewidth}
    {$N_{jet}\geq4$} & 0.93117 & 0.38491 & 0.38328 & 0.10204  \parbox{0pt}{\rule{0pt}{1ex+\baselineskip}}\\ 
     {${N}_{{\emph{b-jet}}} \geq 3$} & 0.24171 & 0.00172 & 0.05706 & 0.02581 \parbox{0pt}{\rule{0pt}{1ex+\baselineskip}}\\ 
     {$N_{\ell\bar{\ell}}\geq1$} & 0.12008 & 0.02146 & 0.09049 & 0.08195 \parbox{0pt}{\rule{0pt}{1ex+\baselineskip}}\\ 
    {$60 \leq {I. M.}_{\, {\ell\bar{\ell}}}  \leq 100$} & 0.07854 & 0.05634 & 0.03264 & 0.03429  \parbox{0pt}{\rule{0pt}{1ex+\baselineskip}}\\ 
    {\textbf{Total eff.}} & \textbf{2.12e-03} & \textbf{8e-07} & \textbf{6e-05} & \textbf{7.4e-06} \parbox{0pt}{\rule{0pt}{1ex+\baselineskip}}\\ \Xhline{3\arrayrulewidth}
   {${\Delta R}{({{ZH}})}\leq1$} & 0.33955 & 0.25000 & 0.61610 & 0.54054 \parbox{0pt}{\rule{0pt}{1ex+\baselineskip}}\\ 
   {\textbf{Total eff.}} & \textbf{7.2e-04} & \textbf{2e-07} & \textbf{4e-05} &  \textbf{4e-06} \parbox{0pt}{\rule{0pt}{1ex+\baselineskip}}\\ \Xhline{3\arrayrulewidth}
  \end{tabular}
\caption{Event selection efficiencies corresponding to background processes. \label{backgroundefftab}}
  \end{center}
\end{table}

\section{Higgs boson reconstruction}
\begin{figure}[h!]
  \centering 
  \bigskip
    \begin{subfigure}[b]{0.62\textwidth} 
    \centering 
     \includegraphics[width=\textwidth]{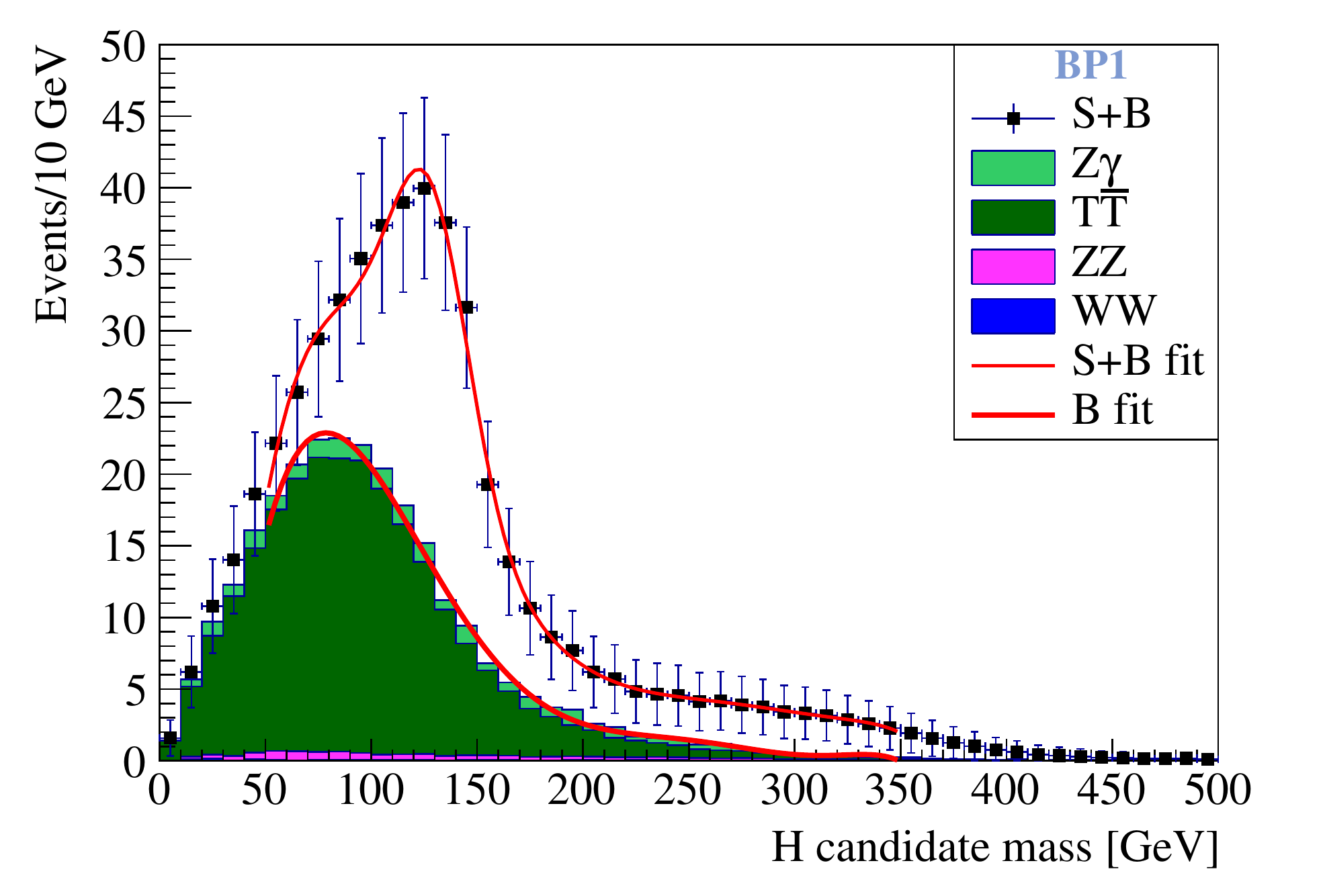}
    \caption{}
    \label{H150Poly}
    \end{subfigure}  
    \bigskip 
    \quad
    \begin{subfigure}[b]{0.62\textwidth} 
    \centering 
    \includegraphics[width=\textwidth]{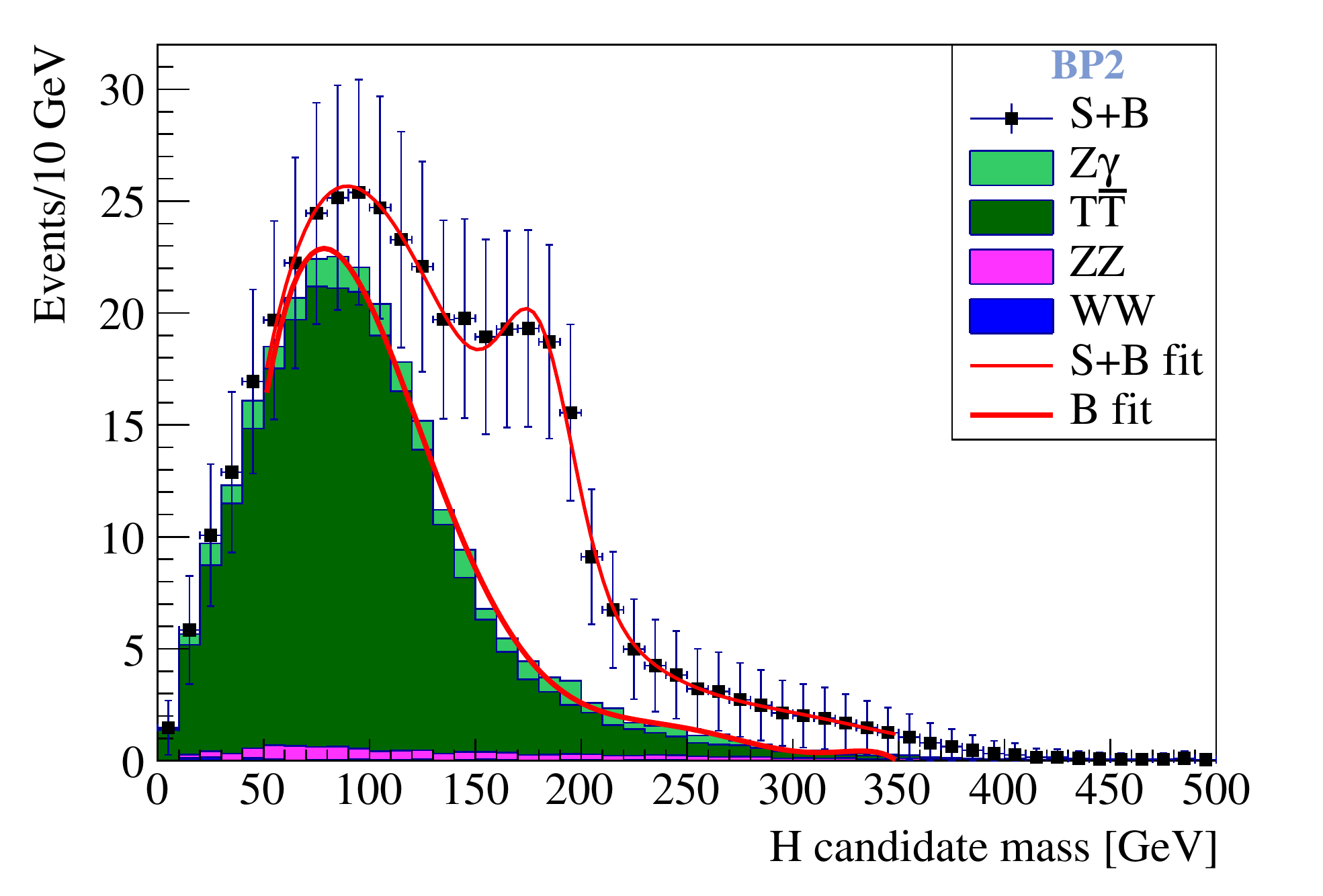}
    \caption{}
    \label{H200Poly} 
    \end{subfigure} 
\caption{$H$ candidate mass distributions with corresponding fitting results assuming the benchmark points a) BP1 and b) BP2 at the integrated luminosity of 500 $fb^{-1}$. Statistical errors are also shown.}
  \label{H12fit}
\end{figure}

\begin{figure}[h!]
   \bigskip
  \centering  
    \begin{subfigure}[b]{0.62\textwidth}
    \centering
    \includegraphics[width=\textwidth]{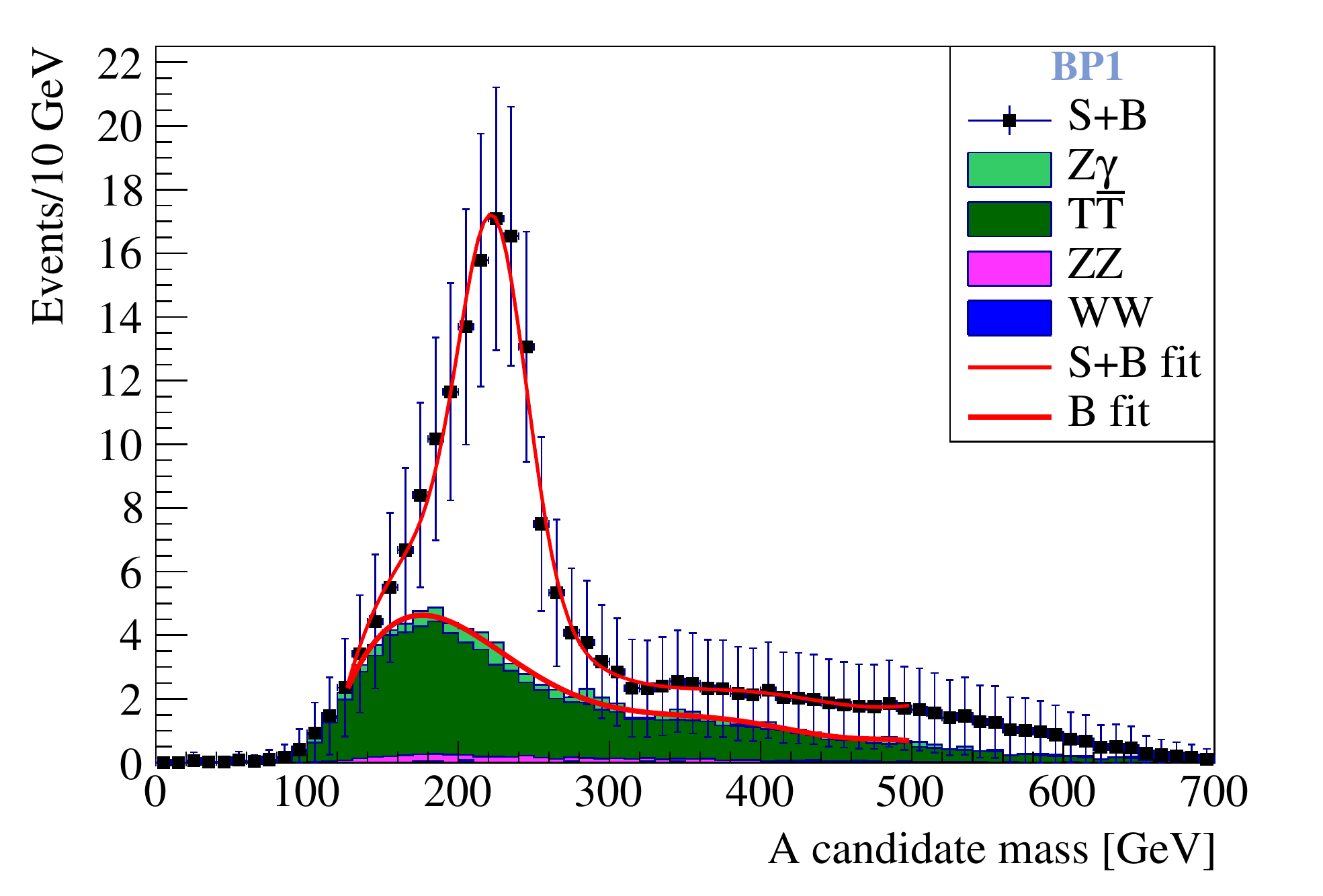}
    \caption{}
    \label{A250Poly}
    \end{subfigure} 
    \bigskip
    \quad    
    \begin{subfigure}[b]{0.62\textwidth}
    \centering
    \includegraphics[width=\textwidth]{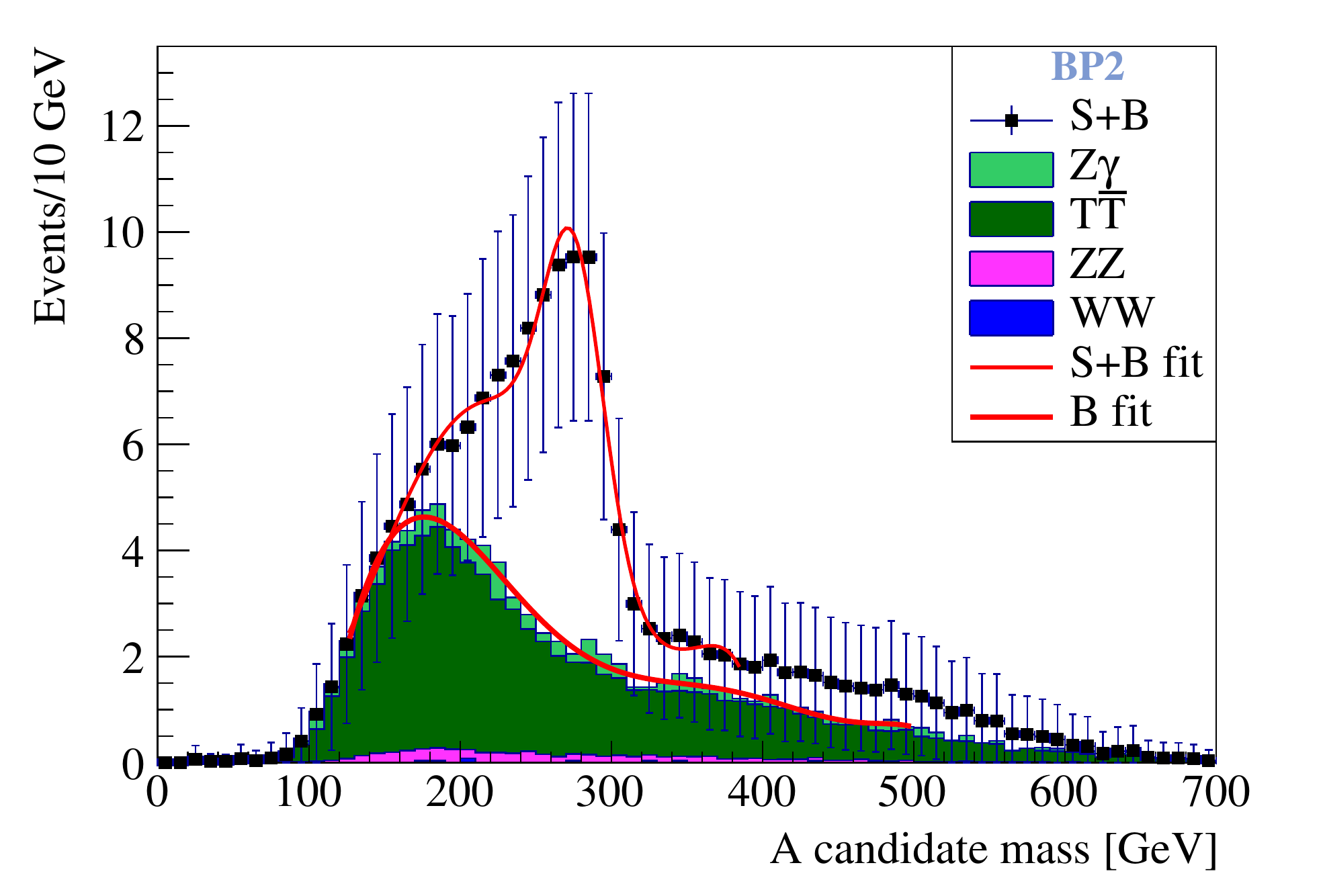}
    \caption{}
    \label{A300Poly}
    \end{subfigure} 
    \caption{$A$ candidate mass distributions with corresponding fitting results assuming the benchmark points a) BP1 and b) BP2 at the integrated luminosity of 500 $fb^{-1}$. Statistical errors are also shown.}
  \label{A12fit}
\end{figure}

Computing candidate masses of the $H$ and $A$ Higgs bosons as explained, mass distributions of Figs. \ref{H12fit} and \ref{A12fit} are obtained. 
Contributions of the signal and different background processes are shown separately and the signal contribution can be seen as a significant excess of data on top of the standard model background. The $WW$ process has the least contribution because of the perfect suppression due to the second and third selection cuts as seen in table \ref{backgroundefftab}. Normalization of the distributions is based on $L\times\sigma\times\epsilon$, where $L$ is the integrated luminosity which is assumed to be 500 $fb^{-1}$, $\sigma$ is the cross section which is taken from tables \ref{sXsec} and \ref{bgXsec}, and $\epsilon$ is the total efficiency. Total efficiencies used for normalizing $A$ mass distributions are taken from the last rows of tables \ref{signalefftab} and \ref{backgroundefftab}. However, total efficiencies corresponding to the first four selection cuts of tables \ref{signalefftab} and \ref{backgroundefftab} cannot be used to normalize $H$ mass distributions since the number of reconstructed $H$ Higgs bosons in events surviving the cuts is different from event to event. Therefore, total efficiencies $\bm{\epsilon_{BP1}}=0.72$ and $\bm{\epsilon_{BP2}}=0.75$ obtained by counting the number of reconstructed $H$ Higgs bosons, are used for normalizing $H$ mass signal distributions and total efficiencies $\bm{\epsilon_{t\bar{t}}}=2.2e-03,\ \ \bm{\epsilon_{WW}}=8.3e-07,\ \ \bm{\epsilon_{ZZ}}=8.8e-05$ and $\bm{\epsilon_{Z\gamma}}=1.1e-05$ are used to normalize background contributions. Benchmark points BP1 and BP2 correspond to Higgs generated masses ${m_H}=150$ and $200$, and ${m_A}=250$ and $300$ GeV respectively. As seen in Figs. \ref{H12fit} and \ref{A12fit}, candidate mass distributions have significant peaks near the generated masses.

Fitting results of the signal plus background and background distributions are also shown in Figs. \ref{H12fit} and \ref{A12fit}. Fitting results are obtained by ROOT 5.34 \cite{root} and the fit function used for the signal plus background distribution is a combination of the polynomial and gaussian functions. A polynomial function is first used as the fit function for the total background distribution and is then used as input for the total signal plus background fit. The Higgs boson reconstructed masses can be read from the ``mean'' parameter of the Gaussian function assumed as the signal distribution function. 

Obtaining values of the ``Mean'' parameter, reconstructed masses of the Higgs bosons $H$ and $A$ are found and provided in table \ref{HAmasstab}.
\begin{table}[h!]
\normalsize
\fontsize{11}{7.2} 
    \begin{center}
         \begin{tabular}{ >{\centering\arraybackslash}m{.18in} >{\centering\arraybackslash}m{1.2in}  >{\centering\arraybackslash}m{.85in}  >{\centering\arraybackslash}m{.85in} >{\centering\arraybackslash}m{.85in} } 
  &  & {BP1} & {BP2} \parbox{0pt}{\rule{0pt}{1ex+\baselineskip}}\\ \Xhline{3\arrayrulewidth}
  \multirow{2}{*}[-3.1pt]{\textbf{H}} &   \cellcolor{blizzardblue}{Gen. mass [GeV]} & 150 & 200  \parbox{0pt}{\rule{0pt}{1ex+\baselineskip}}\\ 
  &   \cellcolor{blizzardblue}{Rec. mass [GeV]} & 127.84$\pm$4.64 & 180.03$\pm$6.51  \parbox{0pt}{\rule{0pt}{1ex+\baselineskip}}\\  \Xhline{2\arrayrulewidth}

 \multirow{2}{*}[-3.1pt]{\textbf{A}} &   \cellcolor{blizzardblue}{Gen. mass [GeV]} & 250 & 300 \parbox{0pt}{\rule{0pt}{1ex+\baselineskip}}\\ 
  &   \cellcolor{blizzardblue}{Rec. mass [GeV]} & 223.24$\pm$4.72 & 275.23$\pm$8.44   \parbox{0pt}{\rule{0pt}{1ex+\baselineskip}}\\ \Xhline{3\arrayrulewidth}
        \end{tabular}
\caption{Reconstructed and generated masses of $H$ and $A$ Higgs bosons with associated uncertainties. \label{HAmasstab}}
  \end{center} 
\end{table} 
\begin{table}[h!]
\normalsize
\fontsize{11}{7.2} 
    \begin{center}
         \begin{tabular}{ >{\centering\arraybackslash}m{.18in} >{\centering\arraybackslash}m{1.5in}  >{\centering\arraybackslash}m{.85in}  >{\centering\arraybackslash}m{.85in} >{\centering\arraybackslash}m{.85in} } 
  &  & {BP1} & {BP2} \parbox{0pt}{\rule{0pt}{1ex+\baselineskip}}\\ \Xhline{3\arrayrulewidth}
  \multirow{2}{*}[-3.1pt]{\textbf{H}} &   \cellcolor{blizzardblue}{Gen. mass [GeV]} & 150 & 200  \parbox{0pt}{\rule{0pt}{1ex+\baselineskip}}\\ 
  &   \cellcolor{blizzardblue}{Corr. rec. mass [GeV]} & 148.91$\pm$10.22 & 201.1$\pm$12.09  \parbox{0pt}{\rule{0pt}{1ex+\baselineskip}}\\  \Xhline{2\arrayrulewidth}

 \multirow{2}{*}[-3.1pt]{\textbf{A}} &   \cellcolor{blizzardblue}{Gen. mass [GeV]} & 250 & 300 \parbox{0pt}{\rule{0pt}{1ex+\baselineskip}}\\ 
  &   \cellcolor{blizzardblue}{Corr. rec. mass [GeV]} & 249.01$\pm$11.3 & 301$\pm$15.02  \parbox{0pt}{\rule{0pt}{1ex+\baselineskip}}\\ \Xhline{3\arrayrulewidth}
        \end{tabular}
\caption{Corrected reconstructed masses of $H$ and $A$ Higgs bosons with associated uncertainties. \label{HAcorrmasstab}}
  \end{center} 
\end{table} 
The difference between generated and reconstructed masses of table \ref{HAmasstab} can be due to the uncertainty arising from fitting method and choice of the fit function, jet reconstruction algorithm and jet mis-identification, $b$-tagging algorithm and jets mis-tag rate and also the errors in energy and momentum of the particles, etc. Optimization of the $b$-tagging algorithm, fitting method and also optimization of the jet reconstruction algorithm using MC truth matching tools can reduce the errors in the reconstructed masses. In addition to the mentioned error sources, electronic noise,  pile up, underlying-events, etc, can give rise to more errors in obtained reconstructed masses in case of real experiments, and thus a careful correction must be applied.

Since the mentioned corrections lie beyond the scope of this paper, a simple off-set correction is applied to the obtained reconstructed masses as follows. On average, reconstructed masses of the $H$ and $A$ Higgs bosons are $21.07$ and $25.77$ GeV smaller than the corresponding generated masses respectively. To apply the off-set correction, reconstructed masses of the $H$ and $A$ Higgs bosons are increased by the same values respectively. Table \ref{HAcorrmasstab} provides the corrected reconstructed masses of the Higgs bosons. The errors are statistical but include also sources of uncertainties from the jet and track energy and momentum resolutions assumed in the analysis. 

\section{Signal significance} 
Using the Higgs candidate mass distributions of Figs. \ref{H12fit} and \ref{A12fit}, signal significance is obtained to assess the observability of the Higgs bosons. A mass window cut is applied to distributions and the signal significance is computed by counting signal and background events which pass the mass window. The mass window is determined by maximizing the signal significance. The integrated luminosity at which computation is performed is set to $500$ $fb^{-1}$. However, it is indicated that the Higgs bosons are also observable at lower luminosities. The minimum required integrated luminosities at which the Higgs bosons are observable with $5\sigma$ signals are also computed and provided as ``$5\sigma$ integrated $\mathcal{L}$.''. Table \ref{sigtab} provides computation results, namely mass window and the corresponding efficiency, signal selection total efficiency, number of signal (S) and background (B) Higgs candidates which pass the mass window, signal to background ratio, signal significance and $5\sigma$ integrated luminosity. 
\begin{table}[h!]
\normalsize
\fontsize{11}{7.2} 
    \begin{center} 
         \begin{tabular}{ >{\centering\arraybackslash}m{.2in} >{\centering\arraybackslash}m{1.50in}  >{\centering\arraybackslash}m{.6in}  >{\centering\arraybackslash}m{.6in} >{\centering\arraybackslash}m{.6in}  } 
  &  & {BP1} & {BP2} \parbox{0pt}{\rule{0pt}{1ex+\baselineskip}}\\ \Xhline{3\arrayrulewidth} 
   &  \cellcolor{blizzardblue}{Gen. mass [GeV]} & 150 & 200 \parbox{0pt}{\rule{0pt}{1ex+\baselineskip}}\\ 
 \multirow{9}{*}[-7.7pt]{\textbf{H}}   & \cellcolor{blizzardblue}{Mass window [GeV]} & >109 & >147  \parbox{0pt}{\rule{0pt}{1ex+\baselineskip}}\\ 
  & \cellcolor{blizzardblue}{Mass window cut eff.} & 0.76 & 0.70 \parbox{0pt}{\rule{0pt}{1ex+\baselineskip}}\\ 
  &  \cellcolor{blizzardblue}{Total eff.} & 0.546 & 0.530  \parbox{0pt}{\rule{0pt}{1ex+\baselineskip}}\\ 
  &  \cellcolor{blizzardblue}{$S$} & 184.7 & 108.9   \parbox{0pt}{\rule{0pt}{1ex+\baselineskip}}\\ 
  &  \cellcolor{blizzardblue}{$B$} & 96.8 & 43.2 \parbox{0pt}{\rule{0pt}{1ex+\baselineskip}}\\  
&\cellcolor{blizzardblue}{$S/B$} & 1.9 & 2.5   \parbox{0pt}{\rule{0pt}{1ex+\baselineskip}}\\  
&\cellcolor{blizzardblue}{$S/\sqrt{B}$} & 18.8 & 16.6   \parbox{0pt}{\rule{0pt}{1ex+\baselineskip}}\\ 
&  \cellcolor{blizzardblue}{Integrated $\mathcal{L}$ [$fb^{-1}]$} &  \multicolumn{2}{c}{500} \parbox{0pt}{\rule{0pt}{1ex+\baselineskip}}\\ 
&  \cellcolor{blizzardblue}{$5\sigma$ integrated $\mathcal{L}$ [$fb^{-1}]$} & 36 & 46  \parbox{0pt}{\rule{0pt}{1ex+\baselineskip}}\\ 
\Xhline{3\arrayrulewidth} 
 &  \cellcolor{blizzardblue}{Gen. mass [GeV]} & 250 & 300  \parbox{0pt}{\rule{0pt}{1ex+\baselineskip}}\\ 
  \multirow{9}{*}[-7.7pt]{\textbf{A}} & \cellcolor{blizzardblue}{Mass window [GeV]} & 191-264 & 232-304  \parbox{0pt}{\rule{0pt}{1ex+\baselineskip}}\\ 
  & \cellcolor{blizzardblue}{Mass window cut eff.} & 0.58 & 0.56  \parbox{0pt}{\rule{0pt}{1ex+\baselineskip}}\\ 
  &  \cellcolor{blizzardblue}{Total eff.} & 0.42 & 0.42  \parbox{0pt}{\rule{0pt}{1ex+\baselineskip}}\\ 
  &  \cellcolor{blizzardblue}{$S$} & 71.2 & 43.6   \parbox{0pt}{\rule{0pt}{1ex+\baselineskip}}\\ 
  &  \cellcolor{blizzardblue}{$B$} & 25.3 & 17.0   \parbox{0pt}{\rule{0pt}{1ex+\baselineskip}}\\  
&\cellcolor{blizzardblue}{$S/B$} & 2.8 & 2.6  \parbox{0pt}{\rule{0pt}{1ex+\baselineskip}}\\ 
&\cellcolor{blizzardblue}{$S/\sqrt{B}$} & 14.2 & 10.6   \parbox{0pt}{\rule{0pt}{1ex+\baselineskip}}\\ 
&  \cellcolor{blizzardblue}{Integrated $\mathcal{L}$ [$fb^{-1}]$} &  \multicolumn{2}{c}{500} \parbox{0pt}{\rule{0pt}{1ex+\baselineskip}}\\
&  \cellcolor{blizzardblue}{$5\sigma$ integrated $\mathcal{L}$ [$fb^{-1}]$} & 63 & 112  \parbox{0pt}{\rule{0pt}{1ex+\baselineskip}}\\ 
\Xhline{3\arrayrulewidth}
        \end{tabular}
\caption{Generated mass, optimized mass window cut and associated efficiency, signal total efficiency, number of signal and background Higgs candidates after all cuts, signal to background ratio, signal significance, assumed integrated luminosity, and the integrated luminosity at which the Higgs boson is observable with a $5\sigma$ signal ($5\sigma$ integrated $\mathcal{L}$).} 
 \label{sigtab}
  \end{center}
\end{table}
According to the results, it is indicated that for both of the benchmark points, both of the Higgs bosons $H$ and $A$ are observable with signals exceeding $5\sigma$ at the integrated luminosity of $500$ $fb^{-1}$ and the minimum required integrated luminosity at which the Higgs boson $H$ ($A$) is observable is 46 (112) $fb^{-1}$. Mass measurement is also possible for both of the Higgs bosons. As seen, minimum required integrated luminosity for observing the heavier Higgs boson is higher. This is because of the fact that cross section of the Higgs production varies inversely as the Higgs mass.

\section{Conclusions} 
The signal process chain $e^- e^+ \rightarrow A H\rightarrow ZHH \rightarrow \ell\bar{\ell} b\bar{b}b\bar{b}$, where $\ell\bar{\ell}$ is a di-electron or a di-muon, was investigated to assess the observability of the CP-even ($H$) and CP-odd ($A$) Higgs bosons in the framework of the Type-\RN{1} 2HDM at SM-like scenario. Electron-positron annihilation is assumed to occur at the center-of-mass energy of $\sqrt{s}=1$ TeV at a linear collider. The signal benefits from large enhancements due to the decay modes $A\rightarrow ZH$ and $H\rightarrow b\bar{b}$ at a relatively low $\tb$ value. The leptonic decay $Z \rightarrow \ell\bar{\ell}$ is assumed to benefit from the clear signature of leptons at lepton colliders. Two benchmark points with different mass hypotheses in the parameter space of the Type-\RN{1} 2HDM were simulated and analysed with the help of characteristics of the signal and background events and appropriate selection cuts. Physical mass of the Higgs boson $H$ ($A$) is assumed to vary in range 150-200 GeV (250-300 GeV). Jet energy smearing was performed according to the energy resolution $\sigma/E=3.5\, \%$. Momentum smearing was also applied to leptons according to the momentum resolution $\sigma_{p_T}/{p_T^2} = 2 \times 10^{-5}$ GeV$^{-1}$. Higgs candidate mass distributions corresponding to assumed benchmark points were obtained and reconstructed masses of the Higgs bosons were obtained by fitting an appropriate fit function to mass distributions and then extracting the value of the ``mean'' parameter of the gaussian fit function. Signal significance was also computed to assess observability of the Higgs bosons. Results indicate that for both of the assumed benchmark points, Higgs bosons $H$ and $A$ are observable with signals exceeding $5\sigma$ at the integrated luminosity of 500 $fb^{-1}$. Mass measurement is also possible for both of the Higgs bosons. Moreover, it was shown that the minimum required integrated luminosities at which the Higgs bosons $H$ and $A$ are observable are 46 and 112 $fb^{-1}$ respectively. Since such luminosities are accessible to future linear colliders, this study is expected to serve experimentalists well in search for Higgs bosons in the context of 2HDM.

\section*{Acknowledgements}
The analysis presented in this work was fully performed using the computing cluster at Shiraz University, college of sciences. We would like to thank Dr. Mogharrab for his careful maintenance and operation of the computing cluster. 

\bibliography{BIB_TO_USE}{}
\bibliographystyle{JHEP}  


\end{document}